\documentclass[11pt, letter]{article}
\usepackage[margin=1in]{geometry}
\usepackage{graphicx, amsmath, amsthm, thmtools, mathtools, todonotes, amsfonts, amssymb,mathrsfs,stmaryrd,enumitem,float,comment,algpseudocode,xspace}
\usepackage{hyperref}
\hypersetup{
     hypertexnames=false
}
\usepackage[capitalise, nameinlink]{cleveref}

\newtheorem{thm}{Theorem}[section]
\newtheorem{clm}[thm]{Claim}

\newtheorem{lem}[thm]{Lemma}
\newtheorem{cor}[thm]{Corollary}

\newtheorem*{que*}{Question}
\theoremstyle{definition}
\newtheorem{defi}[thm]{Definition}
\newtheorem{rem}[thm]{Remark}
\newcommand{\Z}{\mathbb{Z}}

\newcommand{\G}{\mathcal{G}}

\newcommand{\Hi}{\mathcal{H}}
\newcommand{\sm}{\setminus}
\newcommand{\BF}{\texttt{BF}\xspace}

\usepackage[multiuser,inline,nomargin]{fixme}
\fxsetface{inline}{\textmd}
\fxsetup{theme=color}
\FXRegisterAuthor{n}{nsw}{\color{blue}Nicole}
\FXRegisterAuthor{s}{ssh}{Sam}

\title{Dynamic Edge Orientation on Low-Arboricity Graphs:\\ Breaking the $\log n$ Barrier}
\author{Sam Hiken\thanks{\texttt{hiken@umich.edu}}\\University of Michigan \and Nicole Wein\thanks{\texttt{nswein@umich.edu. Supported by NSF CAREER award 2541910. }}\\University of Michigan }
\date{}

\begin{document}

\maketitle

\begin{abstract}

  We study the \emph{dynamic edge orientation problem} introduced in a seminal 1999 paper by Brodal and Fagerberg~\cite{bf99}: given a fully dynamic graph $G$ with \emph{arboricity} bounded by $\alpha$, maintain an orientation on $G$'s edges so every vertex has out-degree upper bounded by a function of $\alpha$. They provided an algorithm with $O(\alpha + \log n)$ amortized update time that maintains an orientation with maximum out-degree $O(\alpha)$. In the decades since, this algorithm has found widespread applications in dynamic graph algorithms. Nevertheless, it has remained a stubborn open question whether, when $\alpha = o(\log n)$, the $O(\log n)$ upper bound can be improved.

    We answer this 27-year-old question in the affirmative, showing that Brodal and Fagerberg's algorithm for maintaining an out-degree $O(\alpha)$ orientation runs in $O(\alpha \cdot \log \log n)$ amortized update time, which is an \emph{exponentially}-improved dependence on $n$. This is the first improvement on Brodal and Fagerberg's upper bound even for the $\alpha=1$ case of dynamic forests. 
    
    Our crucial technical innovation is to construct a hierarchy of dynamic graphs with a property we call \emph{near isolation}. We believe that this construction may be useful for related problems.
\end{abstract}

\pagenumbering{gobble}
\clearpage

\pagebreak
\pagenumbering{arabic}

\section{Introduction}

We study the \emph{dynamic edge orientation problem} introduced in a seminal 1999 paper by Brodal and Fagerberg~\cite{bf99}: given a fully dynamic graph $G$ with \emph{arboricity} bounded by $\alpha$, maintain an orientation on $G$'s edges so every vertex has out-degree upper bounded by a function of $\alpha$.

\emph{Arboricity}\footnote{In this paper we are only concerned with constant-factor approximations of $\alpha$ so we could have equivalently introduced the problem in terms of the graph parameter \emph{degeneracy}, which is within a constant factor of arboricity.}, introduced by Nash-Williams in 1961~\cite{n-w61}, is a widely used measure of the sparsity of a graph that roughly captures the density of the densest subgraph. Formally, an undirected graph $G = (V,E)$ has arboricity $\alpha = \max_{U \subseteq V}\left\lceil \frac{|E(U)|}{|U|-1} \right\rceil$.

Many important graph classes have bounded arboricity, including forests (arboricity 1), planar graphs (arboricity $\le 3$), and graphs excluding fixed minors. Moreover, it is believed that many real-world graphs have low arboricity for their size, such as the Internet and certain social networks~\cite{MR2290727}.

An equivalent characterization of arboricity (and where its name is derived) is given by the Nash-Williams theorem, which says that $G$ has arboricity $\le \alpha$ if and only if $G$ can be decomposed into $\alpha$ edge-disjoint subforests~\cite{n-w61}.

In algorithm design, the most widely exploited property of low-arboricity graphs is that they admit a \emph{low-out-degree edge orientation}. We call an orientation of $G$'s edges a $k$-orientation if every node has out-degree $\le k$. If $G$ has arboricity $\alpha$, then it permits an $\alpha$-orientation. (To see why, take each forest $F_1, ..., F_{\alpha}$ in a forest decomposition of $G$, and for each tree $T$ in $F_i$, orient the edges in $T$ toward a fixed node in $T$.)

In this work, we focus on the dynamic setting, where starting from an empty graph, at each time step an adversary adds or removes an edge from the graph, under the guarantee that the arboricity is always upper bounded by~$\alpha$. Extensive study of dynamic arboricity-$\alpha$ graphs has resulted in algorithms for a wide variety of problems, wherein the most commonly used subroutine is dynamic edge orientation. Thus, it is important to understand the efficiency of this key subroutine. This is the focus of our work.

\subsection{Prior Work on Dynamic Edge Orientation}

In their influential 1999 paper, Brodal and Fagerberg gave a deterministic algorithm for dynamic $O(\alpha)$-orientation with amortized update time $O(\alpha + \log n)$. 

In fact, they prove something stronger: their algorithm, which we refer to as \BF, is constant-competitive with the offline optimal sequence of edge flips.
\begin{lem}[Lemma 1 in \cite{bf99}]\label{lem:bf}
    Let $\G$ be a graph on $n$ nodes undergoing an arboricity $\alpha$-preserving sequence $U$ of edge insertions and deletions, beginning with the empty graph. Let $G_i$ be the state of $\G$ after the $i$-th update.

    Suppose $\Delta > 2\delta \ge 2\alpha$. If there exists a sequence of $\delta$-orientations $\overrightarrow{G}_0, ..., \overrightarrow{G}_{|U|}$ with at most $f$ edge flips in total, then $\BF$ maintains a $\Delta$-orientation of $\G$ and performs at most
    \[ (|U| + f) \frac{\Delta+1}{\Delta+1 - 2\delta}\]
    flips.
\end{lem}

That is, \BF is within a constant factor of optimal in both of the following ways:
\begin{enumerate}
    \item \BF's out-degree $\Delta$ is within a constant factor of the offline optimal out-degree $\delta$, and
    \item Assuming the optimal number of flips is at least amortized constant per edge update, \BF's number of flips is within a constant factor of optimal. 
\end{enumerate}

Furthermore, the total running time of \BF is $O(\alpha|U| + (\#\text{flips }\BF\text{ performs}))$, so it follows from item 2 that \BF's total (and thus amortized) time is within a constant factor of that of the offline optimum. Even though the optimality of \BF (up to constant factors) has long been known, the actual update time of \BF has remained an open problem. 

Although this central problem has remained unresolved, there has been extensive work on various natural extensions. A prominent example is that instead of restricting to out-degree $O(\alpha)$, one can obtain trade-offs between out-degree and update time. Kowalik~\cite{MR2313458} proved that $O(\alpha\log n)$ out-degree in $O(1)$ update time is possible, which is the reverse of $\BF$'s trade-off when restricting to constant $\alpha$. Later, He, Tang, and Zeh obtained a full trade-off between these two extremes, where the two parameters can be any values that multiply to $\log n$, for constant $\alpha$: specifically, for any $\beta\geq 1$ they obtain out-degree $O(\beta\alpha)$ with amortized update time $O\left(\frac{\log (n/(\beta\alpha))}{\beta}\right)$ \cite{htz14}. Those results are for out-degrees larger than that of \BF, but the other side of the trade-off with smaller out-degree, such as $\alpha+1$~\cite{MR4481952} and $\alpha+2$~\cite{MR4473347} have also been studied.

The \BF algorithm is amortized, and it is also interesting to obtain worst-case update time guarantees for dynamic edge orientation. This problem has been studied in several works~\cite{bb18, domingues2026dynamic, kkps14, bkkps21}.

Additionally, while \BF assumes a fixed value of $\alpha$ known in advance, there are dynamic orientation algorithms that adapt to the current value of $\alpha$~\cite{MR4699355,henzinger2020explicit,kkps14,MR4473347, MR4481952}. 

Also, while the out-degree in the \BF algorithm is $O(\alpha)$ after finishing processing each update, the out-degree may blow up temporarily during the flipping process. This issue has been addressed~\cite{ks18}.

Dynamic edge orientation has also been studied for certain subclasses of arboricity-$\alpha$ graphs such as forests~\cite{bkkps21}, planar graphs~\cite{MR4481952}, and outerplanar graphs~\cite{domingues2026dynamic}. 

Edge orientation has also been studied in the related \emph{parallel batch-dynamic} setting~\cite{liu2022parallel, ghaffari2025parallel, blelloch2026faster}. These works also consider the harder, but closely related, problem of maintaining a $k$-core decomposition. 

Additionally, dynamic edge orientation algorithms have been implemented and studied experimentally~\cite{MR4967438,borowitz2023engineering,grossmann2025engineering}.

Beyond the standard dynamic setting, edge orientation has also been considered in other ``local'' settings including the dynamic distributed setting~\cite{MR3478393}, the standard distributed setting~\cite{MR4168052,MR4598820}, the streaming setting~\cite{MR4442076}, and the LCA (local computation algorithms) setting~\cite{MR4802374,levi2026improved}. 

\subsection{Applications of Dynamic Edge Orientation}

Brodal and Fagerberg originally introduced the dynamic edge orientation problem to address the problem of answering adjacency queries. Since then, dynamic edge orientation algorithms have 
been used as a subroutine in arboricity-dependent dynamic algorithms for many problems, including maximal matching~\cite{ns12, htz14, kkps14}, approximate maximum matching~\cite{bs15, bs16, MR3478427}, spanner construction~\cite{bk16}, subgraph counting~\cite{dt12}, approximate densest subgraph~\cite{MR4141752,chekuri20221,MR5082168,MR4870448}, coloring~\cite{sw20,MR4473347,MR4617461,henzinger2020explicit,MR4568163}, maximal independent set~\cite{ossw18}, dominating set~\cite{MR5121598,bosek2026dynamic}, min-cut~\cite{MR5082168}, shortest paths on planar graphs \cite{kk06}, and Cuckoo hashing~\cite{bkkps21}.

Dynamic edge orientation algorithms have even provided useful subroutines for \emph{static} (non-dynamic) problems including coloring triangle-free planar graphs~\cite{k04}, approximating arboricity~\cite{MR4074816}, and approximating the prize-collecting bounded-genus variants of the traveling salesperson problem and Steiner tree~\cite{MR3205243}. 

\subsection{The Main Question}

Despite all of the attention this problem has received, the main open problem raised by Brodal and Fagerberg has remained stubbornly open for 27 years:

\begin{que*} \label{q:main}
On dynamic graphs of arboricity $\alpha =o(\log n)$, is there an algorithm that maintains an $O(\alpha)$-orientation in $o(\log n)$ amortized update time?
\end{que*}

The original amortized update time bound of Brodal and Fagerberg for $O(\alpha)$-orientation has not been improved by \emph{any} super-constant factor since 1999. To make matters worse, there is no non-trivial lower bound; it is possible that an $O(1)$-orientation can be maintained on constant-arboricity graphs in \emph{constant} update time. Moreover, there has been no progress on closing this gap \emph{even for dynamic forests} ($\alpha=1$).

This open problem has been explicitly stated in numerous works~\cite{bf99,bb18,domingues2026dynamic,MR2313458,MR3478427,tang2015fully,berglin2017geometric,bkkps21}, and has been implicit in others.

\subsection{Our Result}
The main contribution of this paper is an improvement over the longstanding $O(\alpha+\log n)$ bound of Brodal and Fagerberg  to $O(\alpha\log\log n)$, an \emph{exponential} improvement in terms of $n$:
\begin{restatable}{thm}{main}\label{thm:main}
    Let $\G$ be a graph on $n$ nodes undergoing an arboricity $\alpha$-preserving sequence $U$ of edge insertions and deletions, beginning with the empty graph. There is a sequence of $(5\alpha-2)$-orientations $\overrightarrow{G}_0, ..., \overrightarrow{G}_{|U|}$ that undergoes at most $O(|U|\cdot \alpha\log \log n)$ edge flips in total.
\end{restatable}
Substituting $\delta = 5\alpha -2$ and $\Delta = 11\alpha$ into \cref{lem:bf} (Brodal and Fagerberg's optimality lemma) yields the following corollary of \cref{thm:main}:
\begin{cor}
     Let $\G$ be a graph on $n$ nodes undergoing an arboricity $\alpha$-preserving sequence $U$ of edge insertions and deletions, beginning with the empty graph. Then the $\BF$ algorithm, with $\Delta = 11\alpha$, performs at most $O(|U| \cdot \alpha \log \log n)$ flips.
\end{cor}
This improves over Brodal and Fagerberg's bound in the regime of $\alpha = o(\log n / \log \log n)$.
\\
\vspace{-.6em}

Brodal and Fagerberg's original motivation was to construct a data structure supporting adjacency queries; we include this application as a corollary:
\begin{cor}
There is a deterministic data structure for answering adjacency queries in a graph undergoing an arboricity $\alpha$-preserving sequence of edge insertions and deletions that has amortized update time $O(\alpha \log \log n)$ and query time $O(\alpha)$.
\end{cor}

Another illustrative application of our result is for dynamic maximal matching. Neiman and Solomon~\cite{ns12} provided a reduction showing that if a $\Delta$-orientation for a graph $\G$ on $n$ vertices under an arboricity $\alpha$-preserving sequence of edge insertions and deletions can be maintained with amortized update time $T$, then a maximal matching
can also be maintained with $O(\Delta+T)$ amortized update time. 

The best-known deterministic algorithm for maintaining a maximal matching on bounded arboricity graphs is given by plugging He, Tang, and Zeh's dynamic edge orientation algorithm into the above reduction to obtain amortized update time $O(\alpha+\sqrt{\alpha\log n})$. Our result implies the following exponential improvement:

\begin{cor}
There is a deterministic data structure for maintaining a maximal matching in a graph $\G$ on $n$ vertices under an arboricity $\alpha$-preserving sequence of edge insertions and deletions that has amortized update time $O(\alpha \log \log n)$.
\end{cor}

It is worth mentioning that on \emph{general graphs}, Solomon provided a randomized algorithm for dynamic maximal matching with \emph{constant} update time, but only against an \emph{oblivious adversary}~\cite{MR3630994}. For randomized algorithms against an adaptive adversary or deterministic algorithms, the best known amortized update time for general graphs is $n^{1/2+o(1)}$~\cite{chuzhoy2026faster}.

Beyond the above corollaries, we expect that our result may open the door to improvements of many of the above variants and extensions of the dynamic edge orientation problem and its associated applications.
The core technical innovation of our paper is the construction of a hierarchy of dynamic graphs with a property we call \emph{near isolation}. This construction may also be a useful tool for related problems.

\subsection{Notation}
We use the term \emph{graph} to mean a multigraph that permits multiple parallel edges but does not permit loops. For a graph $G = (V,E)$, we denote by $G[S]$ the induced subgraph of $G$ on $S \subseteq V$, and $E(S) = E(G[S])$. For any disjoint pair of node sets $S,T \subseteq V$, we use $E(S,T)$ to denote the set of edges with one endpoint in $S$ and one in $T$. If $G$ is directed, then we use $\overrightarrow{E}(S,T)$ to denote the directed edges with tail in $S$ and head in $T$. If $v \in V \sm S$, we use $E(S,v)$ to denote the edges connecting $S$ and $v$ (without regard to direction) and $\overrightarrow{E}(S,v)$ to denote the directed edges from $S$ to $v$.

For a node $v$, let $d^+(v)$ ($d^-(v)$) denote the out-degree (in-degree) of a node $v$, where parallel edges are counted with multiplicity. Let $N^+(v)$ ($N^-(v)$) denote the set of out-neighbors (in-neighbors) of $v$. For any two nodes $u,v$ we denote by $\text{dist}(u,v)$ the distance from $u$ to $v$. We abuse notation in the following way: if $U\subseteq V$, then let $N^{+}(U)$ ($N^-(U)$) denote the set of nodes $v \in V$ so that there exists $u \in U$ such that $(u,v) \in E$ ($(v,u) \in E$). If $U,W \subseteq V$, then let $\text{dist}(U,W) = \min_{u\in U,w \in W}(\text{dist}(u,w))$.

\section{Technical Overview} \label{sub:ov}

In this section, we sketch a proof of \cref{thm:main}.

\main*

We will assume for the purposes of this overview that $\alpha = 1$, so the input is a dynamic forest. The techniques generalize naturally to the larger $\alpha$ case. We will give an offline algorithm for constructing a sequence of $3$-orientations on $G_1, ..., G_{|U|}$ that undergoes $O(|U|\log \log n)$ edge flips in total.

\subsection{Defining the flip sequence}

\paragraph{Dynamic induced subgraph hierarchies.} Given dynamic graph $\G = G_0, ..., G_{|U|}$, a \emph{dynamic induced subgraph} (DIS) of $\G$ is characterized by an ordered pair $(S,I)$, where $S \subseteq V$ and $I = [i,j)$ is an interval $u_i, ..., u_{j-1}$ in the update sequence $U$. We can think of $(S,I)$ as being the dynamic graph defined on the induced subgraphs of $G_{i-1}, G_{i}, ..., G_{j-1}$ on $S$. 

A \emph{DIS hierarchy} $\Hi$ is a binary tree in which each node corresponds to a DIS of $G$. In particular, if $(S,I)$ is an internal node of $\Hi$ with children $(S_1,I_1)$, $(S_2,I_2)$, then the following two properties hold:
\begin{enumerate}
    \item $S_1, S_2 \subseteq S$,
    \item $I = I_1 \sqcup I_2$.
\end{enumerate}

\paragraph{Top-down flip sequence construction.} We assume without loss of generality that $|U| = \text{poly}(n)$ and is a power of 2. Our approach is to construct a DIS hierarchy $\Hi$ on $G$ such that the root of $\Hi$ is $G$ itself (we will describe the details of this construction later). Using $\Hi$, we will construct a sequence of 3-orientations in a top-down fashion, beginning at the root. 

The key to our approach is the following observation: let $(S,I)$ be a DIS of $G$ such that the following two properties hold: 
\begin{enumerate}
    \item If $(u,v)$ is an edge inserted or deleted during $I$, then $u,v \in S$.
    \item We have (somehow) managed to orient every edge in $E(S,V\setminus S)$ toward $S$ at the start of $I$.
\end{enumerate}
If we assume that $G$ is 3-oriented at the start of $I$, then \emph{maintaining} a 3-orientation during the interval $I$ only requires flipping edges with both endpoints in $S$, as no node outside of $S$ can have its out-degree increased.

To this end, let $(S,I) \in V(\Hi)$ with children $(S_1,I_1)$ and $(S_2,I_2)$. In particular, let $I_1 = [i,j)$, $I_2 = [j,k)$, and $I = [i,k)$. A (flawed) approach may look something like this: 
\begin{enumerate}
    \item Orient the edges in $E(S_1,S\setminus S_1)$ toward $S_1$.
    \item Recurse on $S_1$.
    \item Orient the edges in $E(S_2,S\setminus S_2)$ toward $S_2$.
    \item Recurse on $S_2$.
\end{enumerate}
\begin{figure}[t]
    \centering
    \includegraphics[width=0.8\linewidth]{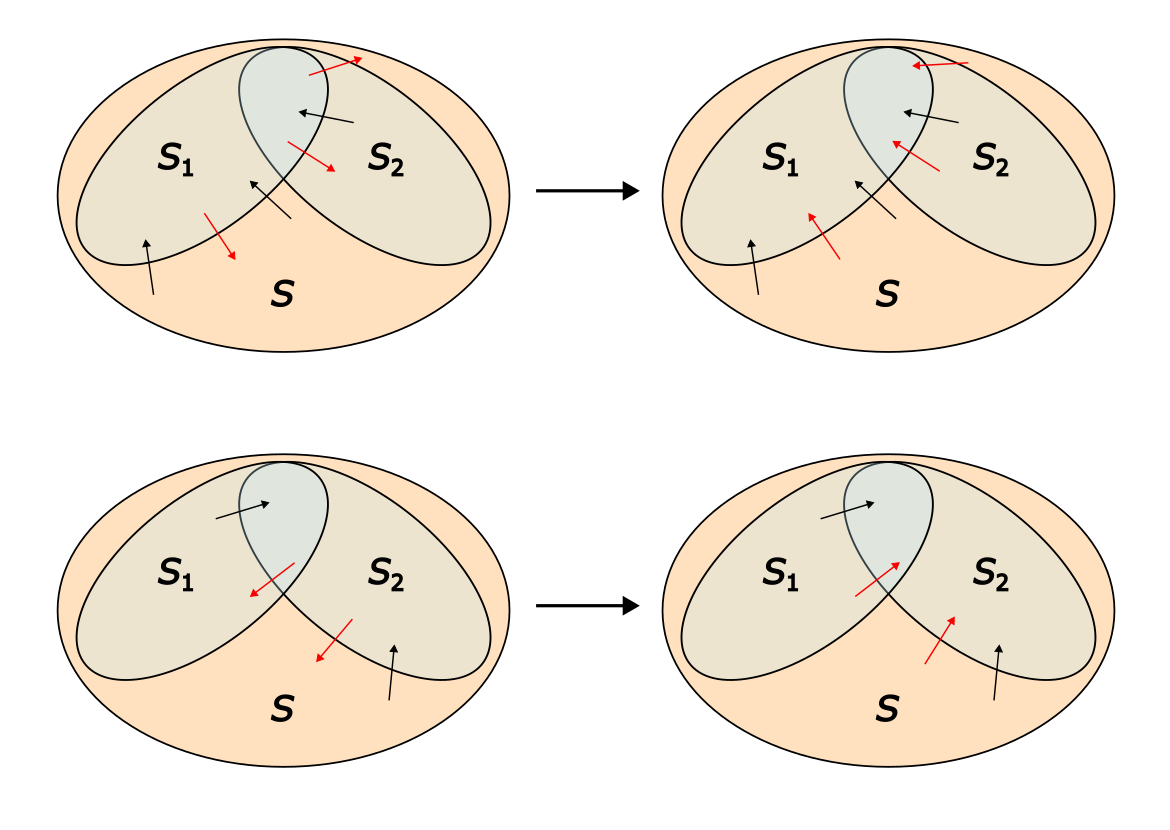}
    \caption{Our algorithm isolates $S_1$ before recursing on $(S_1,I_1)$, and isolates $S_2$ before recursing on $(S_2,I_2)$. This figure illustrates Steps 1 and 3 in the flawed procedure, or Steps 2 and 5 in the correct procedure where we use the batch out-degree reduction lemma.}
    \label{fig:steps1_3}
\end{figure}
Step 1 ensures that every edge in $E(S_1,V \setminus S_1)$ is oriented toward $S_1$ going into update $i$, and hence no additional edge flips outside $E(S_1)$ are needed during $I_1$. Similarly, Step 3 ensures that no edge flips outside of $E(S_2)$ are needed during $I_2$. See \cref{fig:steps1_3} for details.

The problem with this approach is that a node $v$ in $S \setminus S_1$ may have many incident edges in $E(S_1, S \setminus S_1)$, so Step 1 may vastly increase $v$'s out-degree. Similarly, Step 3 may vastly increase the out-degrees of nodes in $S \setminus S_2$. We introduce two technical ingredients to overcome this issue. From here on out, let $W_1 = N^+(S_1) \cap (S \setminus S_1)$ at time step $i$ and $W_2 = N^+(S_2) \cap (S \setminus S_2)$ at time step $j$ (that is, $W_1$ and $W_2$ are the nodes that may have their out-degrees increased by Steps 1 and 3, respectively). Our solution to this problem has two parts.  
\begin{enumerate}
    \item We give a construction of $\Hi$ ensuring that every node in $W_1$ has $\le 1$ neighbor in $S_1$ during $I_1$ (and that, similarly, every node in $W_2$ has $\le 1$ neighbor in $S_2$ during $I_2$). We call this property \emph{near isolation} (\cref{def:near_isolation}).
    \item We add two additional steps to the flip sequence:
    \begin{enumerate}
        \item Before we reorient $E(S_1,S\setminus S_1)$, we ensure that each node in $W_1$ has out-degree $\le 2$.
        \item Before we reorient $E(S_2,S \setminus S_2)$, we ensure that each node in $W_2$ has out-degree $\le 2$.
    \end{enumerate}
\end{enumerate}
This is sufficient to ensure our goal of maintaining a 3-orientation.

\paragraph{Hierarchy construction.} We construct the hierarchy $\Hi$ in a bottom-up fashion, as follows. 

$\Hi$ has $|U|$ leaves (recall that we have asssumed w.l.o.g. that $|U|$ is a power of 2). Let $(u,v)$ be the edge inserted or deleted during the $i$-th update of $U$. The $i$-th leaf node $(S,I)$ will consist of:
\begin{enumerate}
    \item $S = \{ u,v\}$.
    \item $I = [i,i+1)$.
\end{enumerate}
The $\ell$-th level of $\Hi$ will consist of $|U| / 2^{\ell}$ nodes. For nodes $(S_1, I_1)$, $(S_2,I_2)$ on the $(\ell-1)$-th level, construct their parent $(S,I)$ on the $\ell$-th level as follows:
\begin{enumerate}
    \item Let $I = I_1 \sqcup I_2$.
    \item Let $S = S_1 \cup S_2$.
    \item While there is $v \in V \setminus S$ with $\ge 2$ neighbors in $S$ throughout $I$:
    \begin{enumerate}
        \item Add $v$ to $S$.
    \end{enumerate}
\end{enumerate}
\begin{rem}
    Although $\G$ is dynamic, the set of nodes $v \in V \sm S$ with $\ge 2$ neighbors in $S$ considered in Step 3 is time invariant throughout $I$. This is because by induction on the levels of $\Hi$, all edges inserted and deleted during $I_1$ have both endpoints in $S_1$, and all edges inserted and deleted during $I_2$ have both endpoints in $S_2$.
\end{rem}
The above construction ensures that, for every $(S,I) \in V(\Hi)$ with children $(S_1,I_1)$ and $(S_2,I_2)$, each node in $S \setminus S_1$ has $\le 1$ neighbor in $S_1$ throughout $I_1$, and each node in  $S \setminus S_2$ has $\le 1$ neighbor in $S_2$ throughout $I_2$. Thus Steps 1 and 3 in the procedure described above can increase the out-degrees of nodes from 3 to at most 4. We address this issue by first performing a sequence of flips which will ensure the out-degrees of all such nodes are $\le 2$ before we flip the edges along the boundaries of $S_1$ or $S_2$.

\paragraph{Batch out-degree reduction.} Even with our method for constructing $\Hi$, it may be the case that Steps 1 and 3 increment by 1 the out-degrees of nodes in $W_1$ and $W_2$, respectively. In order to ensure that we maintain a 3-orientation, we modify our procedure for constructing a sequence of flips by adding two additional steps (labeled 1 and 4, below):
\begin{enumerate}
    \item Perform a sequence of flips on $E(S\sm S_1)$ so that every node in $W_1$ has out-degree $\le 2$.
    \item Orient the edges in $E(S_1,S\setminus S_1)$ toward $S_1$.
    \item Recurse on $S_1$.
    \item Perform a sequence of flips on $E(S\sm S_2)$ so that every node in $W_2$ has out-degree $\le 2$.
    \item Orient the edges in $E(S_2,S\setminus S_2)$ toward $S_2$.
    \item Recurse on $S_2$.
\end{enumerate}
Now when we orient the edges of $E(S \setminus S_1,S_1)$ and $E(S \setminus S_2,S_2)$ toward $S_1$ and $S_2$, respectively, we can only increase the out-degrees of nodes in $W_1$ and $W_2$ to 3.

We will now provide the idea for bounding the number of flips needed in Step 1 (the bound in Step 4 is similar). A naive bound is $O(|W_1|\log |S \sm S_1|)$, for the following reasoning: let $w \in W_1$. Consider the directed BFS tree $T$ rooted at $w$ restricted to $S \sm S_1$. Because we assumed that $\G$ is a forest, $w$ must be of distance at most $O(\log  |S \sm S_1|)$ to some node $z$ of out-degree $\le 1$ in $S \sm S_1$ (else, $T$ would have depth greater than $\log (|S\sm S_1|)$). We flip every edge in the path from $w$ to $z$. This decrements the out-degree of $w$, increments the out-degree of $z$, and keeps the out-degrees of all intermediate nodes the same. Repeating this sequence of flips for all nodes in $W_1$ yields the $O(|W_1|\log |S \sm S_1|)$ bound. Note that this process does not increase the out-degree of any node to $>2$, and every node in $S \sm S_1$ has $\le 1$ out-edge to $S_1$, so $S$ remains 3-oriented.

It turns out that a better bound is possible. As in the proof of the naive bound, we repeatedly decrement the out-degrees of nodes in $W_1$ by flipping all edges along the path to the nearest node of out-degree $\le 1$. This time, however, we specifically choose to decrement the out-degree of the node in $W_1$ \emph{closest} to a node of out-degree $\le 1$ in $S \sm S_2$. This allows us to take advantage of the following fact: if $W_1$ is ``large,'' then \emph{at least one} node in $W_1$ is close to a node of out-degree $\le 1$.
\begin{enumerate}
    \item Let $W \coloneq W_1$.
    \item While $W \ne \emptyset$:
    \begin{enumerate}
        \item Let $Z$ be the set of nodes in $S \sm S_1$ of out-degree $\le 1$.
        \item Let $P$ be the shortest path from $W$ to $Z$ in $\G[S \sm S_1]$. Call its tail $w$.
        \item Flip every edge in $P$.
        \item Remove $w$ from $W$.
    \end{enumerate}
\end{enumerate}
It turns out that on iteration $i$, the above procedure performs at most 
\[ O \left(\log \left(\frac{|S\sm S_1|}{|W_1|-i+1} \right) \right) \]
flips. Summing over the $|W_1|$ iterations, we perform
\[ O\left( \sum_{i=1}^{|W_1|} \log \left(\frac{|S \sm S_1|}{|W_1|-i+1} \right)\right) = O\left( |W_1| \log \left( \frac{|S \sm S_1|}{|W_1|}\right) \right)\]
flips.

\subsection{Bounding the total number of flips.}\label{subsec:to_bounds}
It turns out that the number of flips that we perform on each subgraph in the hierarchy is bottle-necked by the two calls to the batch out-degree reduction lemma. For any $(S,I) \in \Hi$, let $W_1(I)$ and $W_2(I)$ denote the sets $W_1$ and $W_2$ pertaining to $(S,I)$. The total number of flips performed by our algorithm is:
\[ O\left( \sum_{(S,I) \in V(\Hi)} |W_1(I) \cup W_2(I)|\log \left(\frac{|S|}{|W_1(I) \cup W_2(I)|}\right) \right). \]
In order to bound the above number of flips, it is sufficient to bound the following two quantities:
\begin{enumerate}
    \item $\sum_{(S,I) \in V(\Hi)} |S|$.
    \item $\sum_{(S,I) \in V(\Hi)} |W_1(I) \cup W_2(I)|$.
\end{enumerate}
We will show the following two bounds:
\begin{align}
    \sum_{(S,I) \in V(\Hi)} |S| &= O(|U| \log |U|) \\
    \sum_{(S,I) \in V(\Hi)} |W_1(I) \cup W_2(I)| &= O(|U|).
\end{align}
By the log sum inequality (see \cref{apx:log_sum}), this implies that the total number of flips is 
\[ O\left( \sum_{(S,I) \in \Hi} |W_1(I) \cup W_2(I)|\log \left(\frac{|S|}{|W_1(I) \cup W_2(I)|}\right) \right) = O(|U| \log \log |U|),\]
which is $O(|U| \log \log n)$, as we assumed w.l.o.g. that $|U| = \text{poly}(n)$.

We sketch a proof of Bounds (1) and (2) below.
\paragraph{Proof sketch of (1)} Let $V^{\ell}(\Hi)$ denote the set of DIS's in the $\ell$-th level of $\Hi$, where $V^0(\Hi)$ is the set of leaves. We show that 
\[ \sum_{(S,I) \in V^{\ell}(\Hi)}|S| = O(|U|), \]
for every level $\ell$. By the construction of $\Hi$,
\[ \sum_{(S,I) \in V^{0}(\Hi)}|S| = O(|U|).\]
For $\ell > 0$, we observe a useful fact. For any $(S,I) \in V(\Hi)$ with children $(S_1,I_1)$ and $(S_2,I_2)$, let
\[X(I) = S \setminus (S_1 \cup S_2). \]
In other words, $X(I)$ is the set of nodes added to $S$ not because they appear in $S_1 \cup S_2$, but because they had $\ge 2$ neighbors in $S$ and were iteratively added by the procedure for constructing $\Hi$. Thus, every node contributing to $\sum_{(S,I) \in V^{\ell}(\Hi)}|S|$ either:
\begin{enumerate}
    \item Appears $S_1$ or $S_2$, or
    \item Appears in $X(I)$.
\end{enumerate}
By a short inductive argument, then, every node appearing in some $(S,I) \in V^{\ell}(\Hi)$ either:
\begin{enumerate}
    \item Comes from some $S'$ for $(S',I') \in V^0(\Hi)$, or
    \item Comes from $X(I')$ for some $(S',I')$ which is a descendant of $(S,I)$.
\end{enumerate}
Because we already know that $\sum_{(S,I) \in V^{0}(\Hi)}|S| = O(|U|)$, it suffices to obtain a bound on the second case:
\[ \sum_{(S,I) \in V(\Hi)} |X(I)| = O(|U|).\]
The proof of this fact is given in full detail in \cref{sec:hierarhies}. We observe that for every node $x$ that we include in $X(I)$ for some $(S,I)$, $x$ contributes $\ge 2$ edges to $E(S)$ for the duration of $I$ (the two incident edges that led $x$ to be added to $S$), but only contributes 1 node to $S$ ($x$ itself). Thus, roughly speaking, adding $x$ to $X(I)$ makes $\G[S]$ denser for the duration of $I$. We then show that this can only happen $O(|U|)$ times across $\Hi$ without contradicting the fact that $\G$ is a forest.

\paragraph{Proof sketch of (2)} The next quantity we bound is the total size of the $W_1$'s and $W_2$'s.
\[ \sum_{(S,I) \in V(\Hi)} |W_1(I) \cup W_2(I)| = O(|U|).\]
Here, we discuss a bound on the sum of the $W_1$'s; the bound on the $W_2$'s is similar.
\[ \sum_{(S,I) \in V(\Hi)} |W_1(I)| = O(|U|). \]
Recall that every $w \in W_1(I)$ has $1$ neighbor $u$ in $S_1$. We consider two cases:
\begin{enumerate}
    \item $u \in S_1 \setminus (S_1 \cap S_2)$
    \item $u \in S_1 \cap S_2$.
\end{enumerate}
We handle Case 1 in roughly the same way that we handle the proof that $\sum_{(S,I) \in V(\Hi)}|X(I)| = O(|U|)$. Every edge between $S_1 \setminus (S_1 \cap S_2)$ and $S \setminus S_1$, is an edge that does not appear in $E(S_1) \cup E(S_2)$, but which appears in $E(S)$ for the entire duration of $I$. It turns out that because $\G$ is a forest, this can only happen $O(|U|)$ times across the entire hierarchy.

Next we handle Case 2. For $(S,I)$ with children $(S_1,I_1)$, $(S_2,I_2)$, let $Y(I) = S_1 \cap S_2$. We show the following bound:
\[ \sum_{(S,I) \in V(\Hi)}|Y(I)| = O(|U|).\]
This fact follows from the observation:
\[ \sum_{(S,I) \in V^{\ell}(\Hi)} |S| = \sum_{(S,I) \in V^{\ell-1}(\Hi)} |S| + \sum_{(S,I) \in V^{\ell}(\Hi)}|X(I)| -\sum_{(S,I) \in V^{\ell}(\Hi)}|Y(I)| \]
After rearranging and taking the sum over all levels, we have that:
\[\sum_{(S,I) \in V(\Hi)} |Y(I)| \le \sum_{(S,I) \in V^0(\Hi)} |S| + \sum_{(S,I) \in V(\Hi)} |X(I)|,\]
and we know from the proof of Bound (1) that the RHS is upper bounded by $O(|U|)$.

Now we bound the number of nodes $w \in W_1(I)$ belonging to Case 2. Note that every such $w$ has exactly one in-neighbor in $Y(I)$ before we flip the edges on the boundary of $S_1$. Thus  $w$ is an out-neighbor of some node $y \in Y(I)$. Because every node in $Y(I)$ has out-degree $\le 3$ at the beginning of $I$, the number of nodes in $W_1(I)$ belonging to Case 2 is at most $3 \cdot |Y(I)| = O(|U|)$, as desired.

\subsection{Organization of the paper}
The outline of the rest of the paper is as follows.
\begin{itemize}
	\item In \cref{sec:batch}, we prove a more general version of the batch out-degree reduction lemma (\cref{lem:batch}) that will suit our purposes in the $\alpha > 1$ case.
	\item In \cref{sec:hierarhies}, we describe how we construct our DIS hierarchy $\Hi$. We also prove several bounds on the sizes of node and edge sets in $\Hi$ (in particular, \cref{lem:hier}). These bounds generalize Bounds (1) and (2) from \cref{subsec:to_bounds}.
	\item In \cref{sec:flips}, we construct the sequence of orientations with max out-degree bounded by $O(\alpha)$. We then bound the number of flips needed (\cref{lem:flips}).
\end{itemize}
\section{Batch out-degree reduction} \label{sec:batch}
In this section, we prove the \emph{batch out-degree reduction lemma}, which we use to efficiently reduce the out-degrees of nodes in $S \sm S_1$ and $S \sm S_2$.

\begin{restatable}[Batch out-degree reduction]{lem}{batch}\label{lem:batch}
    Let $H = (V,E)$ be a $\delta$-oriented digraph with arboricity $\le \alpha$, for some $\delta > 2\alpha$. Let $c: V \rightarrow [2\alpha, \delta]$ give a target maximum out-degree of $v \in V$ and let $e(v) = \max\{0,d^{+}(v)-c(v)\}$ be $v$'s excess out-degree. Let $\Phi = \sum_{v \in V}e(v)$. 
    
    There is a sequence of 
    \[O\left(\Phi \cdot \log \left(\frac{(\delta -2\alpha)|V|}{\Phi}\right)\right)\]
    flips such that afterwards each $v \in V$ has out-degree $\le c(v)$.
\end{restatable}
\begin{proof}
    Call $v \in V$ \emph{underfull} if $d^{+}(v) < c(v)$, and \emph{overfull} if $d^{+}(v) > c(v)$. Consider the set of flips performed by the following procedure:
    \begin{enumerate}
        \item While $H$ has an overfull node:
        \begin{enumerate}
            \item Let $W$ be the set of overfull nodes.
            \item Let $Z$ be the set of underfull nodes.
            \item Let $w \in W$ and $z \in Z$ be the nodes that minimize $\text{dist}(w,z)$.
            \item Let $P$ be the shortest $(w,z)$-path in $H$.
            \item Flip every edge in $P$.
        \end{enumerate}
    \end{enumerate}
    Let $H_i$ be the state of $H$ going into the $i$-th iteration, and let $W_i$ and $Z_i$ be the sets $W$ and $Z$ constructed in the $i$-th iteration\footnote{Note that, in the notation $W_i$ here, the subscript does not have the same meaning that $W_1$ and $W_2$ had in the technical overview.}. Let $\Phi_i$ be the value of $\Phi$ going into the $i$-th iteration, so $\Phi_1 = \Phi$.
    
    Observe that each iteration decrements the out-degree of some $w \in W$, increases the out-degree of some underfull node $z$, and keeps the out-degrees of all other nodes the same. Therefore, the procedure terminates after $\Phi$ iterations, and $\Phi_i = \Phi-i+1$. 
    
    We proceed by showing that in the $i$-th iteration, we perform $O\left(\log\left(\frac{(\delta-2\alpha)|V|}{\Phi_i}\right)\right)$ flips. Because the number of flips is determined by the distance from $W_i$ to $Z_i$, this is equivalent to showing
    \begin{align*}
        \text{dist}_{H_i}(W_i,Z_i) = O\left(\log\left(\frac{(\delta-2\alpha)|V|}{\Phi_i}\right)\right).
    \end{align*}
    (Note that distance here is defined as the \emph{directed} distance from $W_i$ to $Z_i$).
    To this end, let
    \[ B_i^r = \{v \in V: \text{dist}_{H_i}(W_i,v) \le r\}\]
    be the ball of radius $r$ around $W_i$ in $H_i$. We claim that for any $i \ge 1$, $r \ge 0$, either
    \begin{enumerate}
        \item $Z_i \cap B_i^r \neq \emptyset$, or 
        \item $2|B_i^r|\le |B_i^{r+1}|$.
    \end{enumerate}
    That is, either $B_i^r$ contains an underfull node, or $B_i^{r+1}$ is at least twice the size of $B_i^r$.

    To see why, suppose that condition 1 does not hold. Then no nodes in $B_i^r$ are underfull, so there are $\ge 2\alpha|B_i^r|$ edges with tails in $B_i^r$, since for all $v\in V$, $c(v)\geq 2\alpha$. Observe that all such edges are contained in $E(B_i^{r+1})$, so
    \[ |E(B_i^{r+1})| \ge 2\alpha|B^r_i|. \]
    Because $H$ has arboricity $\alpha$, 
    \[ |B_i^{r+1}| \ge \frac{1}{\alpha}|E(B_i^{r+1})|\ge \frac{1}{\alpha}2\alpha|B^r_i| = 2|B_i^r|,\]
    as desired. 
    
    Now, let $r'$ be the minimum radius so that $Z_i \cap B_i^{r'} \neq \emptyset$. This is equivalent to saying that
    \[ r' = \text{dist}_{H_i}(W_i,Z_i),\]
    so $r'$ is the number of flips performed in iteration $i$. Clearly, $|B_i^{r'}| \le |V|$. By the above claim, we have that:
    \[ |V| \ge |B_i^0| \cdot 2^{r'},\]
    as the size of the ball doubles each time the radius is incremented. We observe further that $B_i^0 = W_i$, as these are precisely the nodes of distance $0$ from the overfull node set. Therefore, 
    \[ |V| \ge |W_i| \cdot 2^{r'}.\]
    Now, recall from the statement of \cref{lem:batch} that $H$ begins $\delta$-oriented for some $\delta > 2\alpha$, and that every node has target maximum out-degree $\ge 2\alpha$.  Moreover, our procedure does not increase the out-degree of any node beyond $\delta$. Therefore, no individual node in any iteration can have excess greater than $\delta-2\alpha$. Therefore, in iteration $i$ it is the case that
    \[|W_i| \ge  \frac{\Phi_i}{\delta-2\alpha},\]
    as the nodes in $W_i$ are exactly the nodes contributing to $\Phi_i$. Putting everything together, we have that
    \[ |V| \ge |W_i| \cdot 2^{r'} \ge  \frac{\Phi_i}{\delta-2\alpha} \cdot 2^{r'}, \]
    and rearranging we get that
    \[ r' \le \log_2 \left( \frac{(\delta-2\alpha)|V|}{\Phi_i} \right), \]
    where the RHS bounds the number of flips performed in iteration $i$.

    It remains to sum the number of flips needed across all rounds. We know that:
    \begin{align*}
        \text{total } \# \text{flips} &=O\left( \sum_{i = 1}^{\Phi} \log\left(\frac{(\delta-2\alpha)|V|}{\Phi_i}\right)\right)\\
        &= O\left(\sum_{i = 1}^{\Phi} \log\left(\frac{(\delta-2\alpha)|V|}{\Phi - i+1}\right)\right)\\
        &= O\left(\sum_{i = 1}^{\Phi}\left(\log\left((\delta-2\alpha)|V|\right) - \log\left(\Phi - i+1\right)\right)\right)\\
        &= O\left(\Phi\log\left((\delta-2\alpha)|V|\right) - \sum_{i = 1}^{\Phi}\log\left(\Phi - i+1\right)\right) \\
        &= O\left(  \Phi\log\left((\delta-2\alpha)|V|\right) - \int_1^{\Phi}\log( x)\space dx\right) \\
        &= O\left(\Phi\log\left((\delta-2\alpha)|V|\right) - \Phi\log(\Phi) \right)\\
        &= O\left(\Phi\log\left(\frac{(\delta-2\alpha)|V|}{\Phi}\right) \right),
    \end{align*}
    as desired.
   
\end{proof}
\section{DIS hierarchies}\label{sec:hierarhies}

In this section, we describe an offline procedure for constructing from $\G$ a hierarchy of dynamic subgraphs. We will ultimately use this hierarchy to construct the flip sequence used in the proof of \cref{thm:main}.

\begin{defi}[Dynamic Induced Subgraph]
    Given a dynamic (undirected) graph $\G$ on node set $V$ undergoing updates $U$, a \emph{dynamic induced subgraph} (DIS) is the dynamic graph characterized by $(S,I)$, where $S \subseteq V$ and $I = [i,j)$ is an interval of updates $u_i, ... u_j$.
\end{defi}

\begin{rem}
    Throughout this paper, we think of update $u_i$ as transitioning $\G$ from $G_{i-1}$ to $G_i$.
\end{rem}

\begin{defi}[Total DIS]
    A \emph{total} DIS $(S,I)$ of $\G$ is a DIS in which every edge inserted and deleted in $\G$ during $I$ has both endpoints in $S$.
\end{defi}

\begin{defi}[Near Isolation]\label{def:near_isolation}
    A \emph{nearly isolated} DIS $(S,I)$ of $\G$ is a DIS in which no node $v \in V \sm S$ has $|E(S,v)| \ge 2\alpha$ any point during $I$. 
\end{defi}

\begin{defi}[DIS Hierarchy]\label{def:dis_h}
    A \emph{DIS hierarchy} $\Hi$ on $\G$ is a binary tree with a DIS of $\G$ at each node such that, 
    \begin{enumerate}
        \item if $(S,I)$ is a leaf of $\Hi$, then $I = [i,i+1)$ for some $i$, and
        \item if $(S,I)$ has children $(S_1, I_1)$ and $(S_2, I_2)$, then
        \begin{enumerate}
            \item $S_1\cup S_2 \subseteq S$, and
            \item $I_1 = [i,j)$, $I_2 = [j,k)$, and $I = [i,k)$ for some $i<j<k$.
        \end{enumerate}
    \end{enumerate}
\end{defi}

\begin{defi}[Special DIS Hierarchy]
    A DIS hierarchy $\Hi$ is \emph{special} if every DIS in $\Hi$ is total and nearly isolated.
\end{defi}

\begin{defi}
    Let $\Hi$ be a DIS hierarchy.
    \begin{itemize}
        \item Let $V^{leaf} (\Hi)$ be the set of DIS's at leaves of $\Hi$
        \item Let $V^{int}(\Hi)$ be the set of DIS's at internal nodes of $\Hi$
        \item Let $V^{\ell} (\Hi)$ be the set of DIS's at level $\ell$ in $\Hi$ (with the leaves forming level 0).
    \end{itemize}
\end{defi}

\begin{defi}
    Fix a DIS hierarchy $\Hi$. For any $(S,I) \in V^{int}(\Hi)$ with children $(S_1, I_1)$, $(S_2,I_2)$, let
    \begin{itemize}
        \item $X(I) \coloneq S \sm (S_1 \cup S_2)$
        \item $Y(I) \coloneq S_1 \cap S_2$
        \item $A(I) \coloneq S_1 \sm Y(I)$
        \item $B(I) \coloneq S_2 \sm Y(I)$.
    \end{itemize}
\end{defi}
\begin{figure}[t]
    \centering
    \includegraphics[width=0.5\linewidth]{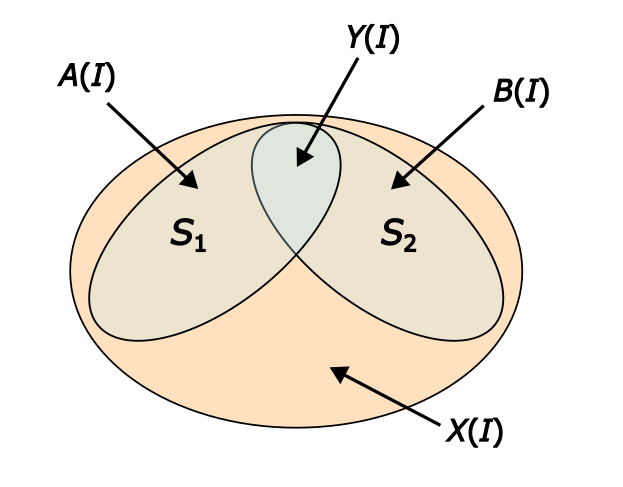}
    \caption{The sets $A(I)$, $B(I)$, $X(I)$, and $Y(I)$.}
    \label{fig:abxy}
\end{figure}
See \cref{fig:abxy} for an illustration of the sets $X(I)$, $Y(I)$, $A(I)$, and $B(I)$ in $S$.

The remainder of this section is devoted to proving the following lemma. 
\begin{lem} \label{lem:hier}
    Let $\G = G_0, ..., G_{|U|}$ be a dynamic graph on node set $V$ undergoing an arboricity $\alpha$-preserving sequence $U = \{ u_1, ..., u_{|U|}\}$ of edge insertions and deletions. $G_0$ is not necessarily the empty graph. Suppose that $|U|$ is a power of 2. There is a special DIS hierarchy $\Hi$ on $\G$ such that:
    \begin{enumerate}
        \item $\sum_{(S,I) \in V^{int}(\Hi)} |X(I)| = O(|U|)$
        \item $\sum_{(S,I) \in V^{int}(\Hi)} |Y(I)| = O(|U|)$
        \item $\sum_{(S,I) \in V(\Hi)} |S| = O(|U|\log|U|)$
        \item $\sum_{(S,I) \in V^{int}(\Hi)} |E(A(I),B(I))|=O( \alpha|U|)$,
    \end{enumerate}
    where $|E(A(I),B(I))|$ denotes the number of edges crossing $A(I)$,$B(I)$.
\end{lem}

\begin{rem}
    Although $\G$ is dynamic, $|E(A(I),B(I))|$ remains constant throughout the interval $I$. This is because $\Hi$ is assumed to be total, so all edges inserted or deleted during $I_1$ have both endpoints in $S_1$, and all edges inserted or deleted during $I_2$ have both endpoints in $S_2$.
\end{rem}

\begin{proof}[Proof of \cref{lem:hier}]
    Construct $\Hi$ as follows:
\begin{enumerate}
    \item For each $u_i \in U$, let $(w,v)$ be the edge that is deleted or inserted in $u_i$. Add $(\{ w,v\}, [i, i+1))$ to $V^{leaf}(\Hi)$.
    \item For $\ell = 1, ..., \log_2|U|$:
    \begin{enumerate}
        \item For $q = 0, ..., |U|/2^{\ell}-1:$
        \begin{enumerate}
            \item Let $i = q\cdot 2^{\ell}+1$,
            $j = (q+\frac{1}{2}) \cdot 2^{\ell}+1$, and $k = (q+1)\cdot 2^{\ell}+1$.
            \item Let $(S_1,I_1) \in V(\Hi)$ be the DIS with $I_1 = [i,j)$, and let $(S_2, I_2) \in V(\Hi)$ be the DIS with $I_2 = [j,k)$.
            \item $I \coloneq [i,k)$.
            \item $S \coloneq S_1 \cup S_2 $.
            \item While there is $v \in V \sm S$ with $|E(v, S)| \ge 2\alpha$: add $v$ to $S$.
            \item Add $(S,I)$ to $V^{int}(\Hi)$, with children $(S_1, I_1)$, $(S_2, I_2)$.
        \end{enumerate}
    \end{enumerate}
\end{enumerate}
Step v. is well-defined because $S_1$ is total during $I_1$ and $S_2$ is total during $S_2$, so for any node $v \in V \sm S$, the edge set $E(v, S)$ is time invariant throughout $I$.

It is straightforward to see that $\Hi$ is a DIS hierarchy under \cref{def:dis_h}. We first show that $\Hi$ is special, then prove that it obeys the four requisite bounds.
\begin{clm}
    $\Hi$ is special.
\end{clm}
\begin{proof}
    Fix $(S,I) \in V(\Hi)$. We will show that $(S,I)$ is total and nearly isolated by induction on $(S,I)$'s level in $\Hi$. 

    In the base case, $(S,I = [i, i+1)) \in V^{leaf}(\Hi)$, and Step (1) guarantees that $(S,I)$ is total. To see that $(S,I)$ is nearly isolated, suppose for contradiction there is a node $x \in V \sm S$ with $|E(S, x)|\ge2\alpha$ before and after update $i$. Let $v,w$ be the nodes of $S$. If $i$ is an insertion, then $G_i[{v,w,x}]$ contains $2\alpha+1$ edges. This implies the arboricity of $G_i$ is at least $(2\alpha+1)/2>\alpha$, a contradiction. If $i$ is a deletion, the argument is symmetric by considering $G_{i-1}$.
    
    In the inductive case, $(S,I)$ has children $(S_1, I_1)$ and $(S_2,I_2)$ that are total and near isolated. Because $I = I_1 \sqcup I_2$ and $S_1 \cup S_2 \subseteq S$, $(S,I)$ is total. Moreover, $(S,I)$ is nearly isolated, as any node with $\ge 2\alpha$ neighbors in $S$ is added to $S$ by step (v).
\end{proof}
It remains to show that $\Hi$ obeys the bounds in \cref{lem:hier}. The following two inequalities are an important step in obtaining these bounds.
\begin{clm} \label{clm:ptl}
    For any $(S,I = [i,k)) \in V^{int}(\Hi)$ with children $(S_1,I_1 = [i,j))$, $(S_2,I_2 = [j,k))$, the following two inequalities hold:
    \begin{align}
        |E_{G_{i-1}}(S)| &\ge |E_{G_{i-1}}(S_1)|+|E_{G_{j-1}}(S_2)| + |E(A(I),B(I))| + 2\alpha|X(I)| - \alpha|Y(I)| \\
        |E_{G_{k-1}}(S)| &\ge |E_{G_{j-1}}(S_1)|+|E_{G_{k-1}}[S_2])| + |E(A(I),B(I))| + 2\alpha|X(I)| - \alpha|Y(I)|
    \end{align}
\end{clm}
Note that because $E(A(I),B(I))$ is invariant throughout $I$, we do not specify a time step $i$, $j$, or $k$ as we do with $E(S)$, $E(S_1)$, and $E(S_2)$.
\begin{proof}
    We will prove (1). The proof of (2) is symmetric, replacing $S_1$ with $S_2$ and $i$ with $k$. Observe that an edge in $E_{G_{i-1}}(S)$ is of one of four types:
    \begin{enumerate}
        \item Both endpoints are in $S_1$,
        \item One endpoint is in $A(I)$, and one endpoint is in $B(I)$,
        \item Both endpoints are in $S_2$, at most one endpoint is in $S_1$,
        \item At least one endpoint is in $X(I)$.
    \end{enumerate}
    Bound (1) in \cref{clm:ptl} follows from summing the following four (in)equalities. 1 and 2 are trivial. We provide proofs of 3 and 4:
    \begin{enumerate}
        \item $|E_{G_{i-1}}(S_1)| = \#\text{ edges of type 1}$,
        \item $|E(A(I),B(I))| = \#\text{ edges of type 2}$,
        \item $|E_{G_{j-1}}(S_2)| - \alpha|Y(I)| \le \#\text{ edges of type 3}$
        \begin{proof}
            Because $G_{j-1}$ has arboricity $\le \alpha$, $|E_{G_{j-1}}(Y(I))| \le \alpha |Y(I)|$. Therefore, \[|E_{G_{j-1}}(S_2)| - \alpha|Y(I)| \le (\#\text{ edges in }G_{j-1}[S_2]\text{ with} \le 1\text{ endpoint in }Y(I)).\]
            Because $(S_1,I_1)$ is total, all edges added during $I_1$ have both endpoints in $S_1$. Therefore, none of the edges in $G_{j-1}[S_2]$ with $\le 1$ endpoint in $Y(I)$ are added during $I_1$, so they all exist at time step $i$. Therefore, 
            \[|E_{G_{j-1}}(S_2)| - \alpha|Y(I)| \le (\#\text{ edges in }G_{i-1}[S_2]\text{ with} \le 1\text{ endpoint in }Y(I)),\]
            where the right-hand side is equivalent to the edges of type 3. 
        \end{proof}
        \item $2\alpha|X(I)| \le \#\text{ edges of type 4}$.
        \begin{proof}
            Consider the while-loop in step v. of the procedure to construct $\Hi$. A node $v$ is added to $S$ only if $|E(v,S)| \ge 2\alpha$ throughout $I$. Observe that any edge counting toward $|E(v,S)|$ here is an edge of type 4. The inequality follows.
        \end{proof}
    \end{enumerate}
    This concludes the proof of \cref{clm:ptl}.
\end{proof}
An important ingredient in the rest of the proof of \cref{lem:hier} is a potential function, $\Psi$, defined on the levels of $\Hi$. For $(S,I = [i,k)) \in  V(\Hi)$, let 
    \[\psi(I) \coloneq 2\alpha|S| - |E_{G_{i-1}}(S)| - |E_{G_{k-1}}(S)|.\] 
Note that $\alpha|S|$ is an upper bound on the number of edges in $\G[S]$ at a particular time step, so for intuition we can think of $\psi(I)$ as measuring the sum of edges ``missing'' from $S$ at the beginning of $I$ and the number ``missing'' at the end of $I$. In particular, when $\alpha = 1$, $\psi(I)$ is measuring the number of connected components in $\G[S]$ a the start of $I$ $+$ the number of connected components in $\G[S]$ at the end of $I$.
    
For a level $\ell \in [0, \log_2|U|]$, let
    \[\Psi(\ell) \coloneq \sum_{(S,I) \in V^{\ell}(\Hi)}\psi(I).\]
\begin{clm}\label{clm:x_ab}
    For any level $\ell \ge 1$, \[\Psi(\ell-1) -\Psi(\ell) \ge 2\sum_{(S,I) \in V^{\ell}(\Hi)} \left(\alpha|X(I)| + |E(A(I),B(I))| \right).\]
\end{clm}
\cref{clm:x_ab} is stating that as we go up the tree, at each level the potential decreases by at least the quantity on the RHS. Intuitively, this is because the RHS is measuring the number of edges introduced to the DIS's $(S,I)$ on level $\ell$ that did not appear in children $(S_1,I_1)$ or $(S_2,I_2)$. In the example of $\alpha=1$, we can think of these edges as joining up components that on the level below were disconnected for the duration of $I_1$ and $I_2$.

\begin{proof}
    Let $(S,I) \in V^{\ell}(\Hi)$ with children $(S_1,I_1)$, $(S_2,I_2)$. By the definition of $\psi$,
    \begin{align*}
        \psi(I) &= 2\alpha|S| - |E_{G_{i-1}}(S)| - |E_{G_{k-1}}(S)|. 
    \end{align*}
    Plugging in the inequalities from \cref{clm:ptl},
    \begin{align*}
        \psi(I) \le 2\alpha|S| &-|E_{G_{i-1}}(S_1)|-|E_{G_{j-1}}(S_2)|- |E(A(I),B(I))|- 2\alpha|X(I)|+ \alpha|Y(I)| \\
        &-|E_{G_{j-1}}(S_1)|-|E_{G_{k-1}}(S_2)|- |E(A(I),B(I))|- 2\alpha|X(I)|+ \alpha|Y(I)|.
    \end{align*}
    Because $|S| + |Y(I)|= |S_1| + |S_2| + |X(I)|$,
    \begin{align*}
        \psi(I) \le 2&\alpha|S_1| -|E_{G_{i-1}}(S_1)|-|E_{G_{j-1}}(S_2)|- |E(A(I),B(I))|- \alpha|X(I)|\\
        +&2\alpha|S_2|-|E_{G_{j-1}}(S_1)|-|E_{G_{k-1}}(S_2)|- |E(A(I),B(I))|- \alpha|X(I)|.
    \end{align*}
    Plugging in the definition of $\psi$,
    \begin{align*}
        \psi(I) &\le \psi(I_1)- |E(A(I),B(I))|- \alpha|X(I)|+\psi(I_2)- |E(A(I),B(I))|- \alpha|X(I)| \\
        &= \psi(I_1) + \psi(I_2)  - 2\alpha|X(I)| - 2|E(A(I),B(I))|.
    \end{align*}
    Rearranging, we get
    \[ \psi(I_1) + \psi(I_2) - \psi(I) \ge 2\alpha|X(I)| +2|E(A(I),B(I))|,\]
    and \cref{clm:x_ab} follows from summing over all DIS's in $V^{\ell}(\Hi)$.
\end{proof}
\cref{clm:x_ab} provides a bound on the change in $\Psi$ between layers. We will now use $\Psi$ as a potential function to bound the LHS in the expression below.

\begin{clm}\label{clm:x_ab_sum}
\[\sum_{(S,I) \in V^{int}(\Hi)}\alpha|X(I)| + |E(A(I),B(I))| = O(\alpha|U|).\]
\end{clm}
\begin{proof}
    Because a leaf DIS $(S,[i,i+1))$ contains exactly two nodes that share an edge in either $G_i$ or $G_{i+1}$, 
    \[ \psi([i,i+1)) \le 4\alpha-1.\]
    Taking the sum over all leaves,
    \[ \Psi(0) \le |U|(4\alpha - 1). \]
    Summing both sides of \cref{clm:x_ab} across all internal levels of $\Hi$ we get that
    \begin{align*}
        2 \sum_{(S,I) \in V^{int}(\Hi)}\alpha|X(I)| + |E(A(I),B(I))| &\le \Psi(0) \\
        &\le |U|(4\alpha-1),
    \end{align*}
    and \cref{clm:x_ab_sum} follows.
\end{proof}
\begin{cor}[Bound 1. in \cref{lem:hier}] \label{cor:x_bound}
    \[\sum_{(S,I) \in V^{int}(\Hi)} |X(I)| = O(|U|).\]
\end{cor}
\begin{cor}[Bound 4. in \cref{lem:hier}] \label{cor:ab_bound}
    \[\sum_{(S,I) \in V^{int}(\Hi)} |E(A(I),B(I))| = O(\alpha|U|).\]
\end{cor}
\begin{clm}[Bound 2 in \cref{lem:hier}]
    \[\sum_{(S,I) \in V^{int}(\Hi)} |Y(I)| = O(|U|).\]
\end{clm}
\begin{proof}
    First, observe that for any $(S,I) \in V^{int}(\Hi)$ with children $(S_1,I_1)$, $(S_2,I_2)$, 
    \[|Y(I)| = |S_1| + |S_2| -|S| +|X(I)|. \]
    Summing over all DIS's in an internal level $\ell$,
    \[\sum_{(S,I) \in V^{\ell}(\Hi)} |Y(I)| = \sum_{(S,I) \in V^{\ell-1}(\Hi)}|S| - \sum_{(S,I) \in V^{\ell}(\Hi)}|S| + \sum_{(S,I) \in V^{\ell}(\Hi)} |X(I)|.\]
    Summing across all internal levels, we get a telescoping sum that evaluates to
    \begin{align*}
        \sum_{(S,I) \in V^{int}(\Hi)} |Y(I)| &\le \sum_{(S,I) \in V^{leaf}(\Hi)}|S| + \sum_{(S,I) \in V^{int}(\Hi)} |X(I)|.
    \end{align*}
    Because each leaf DIS has two nodes, 
    \[\sum_{(S,I) \in V^{leaf}(\Hi)}|S| = 2|U|,\]
    and by \cref{cor:x_bound}, 
    \[\sum_{(S,I) \in V^{int}(\Hi)} |X(I)| = O(|U|).\] 
    It follows that
    \[ \sum_{(S,I) \in V^{int}(\Hi)} |Y(I)|  = O(|U|), \]
    as desired.
\end{proof}
\begin{clm}[Bound 3 in \cref{lem:hier}]
    \[\sum_{(S,I) \in V(\Hi)} |S|=O( |U|\log|U|).\]
\end{clm}
\begin{proof}
    There are $\log_2(|U|)+1$ levels in $\Hi$, so we will show that for each level $\ell$,
    \[ \sum_{(S,I) \in V^{\ell}(\Hi)} |S| = O(|U|).\]
    Note that for any $(S,I) \in V^{int}(\Hi)$ with children $(S_1,I_1)$, $(S_2,I_2)$, every node in $S$ is either in $S_1$, $S_2$, or $X(I)$. Therefore, for $\ell \ge 1$,
    \[ \sum_{(S,I) \in V^{\ell}(\Hi)} |S| \le \sum_{(S,I) \in V^{\ell-1}(\Hi)} |S| + \sum_{(S,I) \in V^{\ell}(\Hi)}|X(I)|.\]
    Solving the recurrence relation, we get that 
    \begin{align*}
        \sum_{(S,I) \in V^{\ell}(\Hi)} |S| &\le \sum_{(S,I) \in V^{leaf}(\Hi)} |S| + \sum_{j = 1}^{\ell} \left(\sum_{(S,I) \in V^{j}(\Hi)}|X(I)| \right) \\
        &\le \sum_{(S,I) \in V^{leaf}(\Hi)} |S| + \sum_{(S,I) \in V^{int}(\Hi)} |X(I)| \\
        &= O(|U|),
    \end{align*}
    with the last bound using \cref{cor:x_bound}.
\end{proof}
This concludes the proof of \cref{lem:hier}. 
\end{proof}

\section{Constructing the flip sequence} \label{sec:flips}
In this section, we use the decomposition from \cref{sec:hierarhies} to construct (offline) a sequence of flips on $\G$ that matches the bound of \cref{thm:main}.

The following notation will be useful:
\begin{defi}\label{def:gamma}
    Let $(S,I) \in V^{int}(\Hi)$ for a special hierarchy $\Hi$. Let
    \[ \Gamma(I) = \alpha |X(I)| +\alpha|Y(I)| + |E(A(I),B(I))|\]
\end{defi}
The bulk of this section will be spent proving the following core lemma.
\begin{lem} \label{lem:flips}
    Let $\G = G_0, ..., G_{|U|}$, $U$, and $\Hi$ be as described in \cref{lem:hier}. Let $(S,I= [i,k)) \in V(\Hi)$.
\begin{enumerate}
    \item If $(S,I) \in V^{leaf}(\Hi)$, then for any initial $(5\alpha-2)$-orientation on $G_{i-1}[S]$, there is a $(5\alpha-2)$-orientation on $G_{k =i+1}[S]$ that does not require any edge reorientations.
    \item If $(S,I) \in V^{int}(\Hi)$, then let $desc(S,I)$ denote the set of DIS's that are descendants of $(S,I)$ in $V^{int}(\Hi)$ (including $(S,I)$ itself). For any initial $(5\alpha-2)$-orientation on $G_{i-1}[S]$, there exists a sequence of $(5\alpha-2)$-orientations on $G_{i}[S], ..., G_{k-1}[S]$ such that the total number of edge reorientations is less than
    \[ O\left( \sum_{(S',I') \in desc(S,I)} \Gamma(I')\log\left(\frac{\alpha|S'|}{\Gamma(I')}\right)\right). \]
\end{enumerate}
\end{lem}
\begin{proof}
    We proceed by induction on the level of $(S,I)$. In the base case, $(S,I) \in V^{leaf}(\Hi)$, so $|S| = 2$, so $|E(G_{i-1}[S])|, |E(G_{i}[S])| \le \alpha$, so we do not need to perform any flips to maintain a $(5\alpha-2)$-orientation. 
    
    In the inductive case, we construct the $(5\alpha-2)$-orientations as follows. 
    Let $(S_1, I_1 = [i,j))$, $(S_2,I_2 = [j,k))$ be the children of $(S,I)$. Construct the orientations $\overrightarrow{G}_{i}, ..., \overrightarrow{G}_{k-1}$ 
    as follows:
    \begin{figure}[t]
    \centering
    \includegraphics[width=0.8\linewidth]{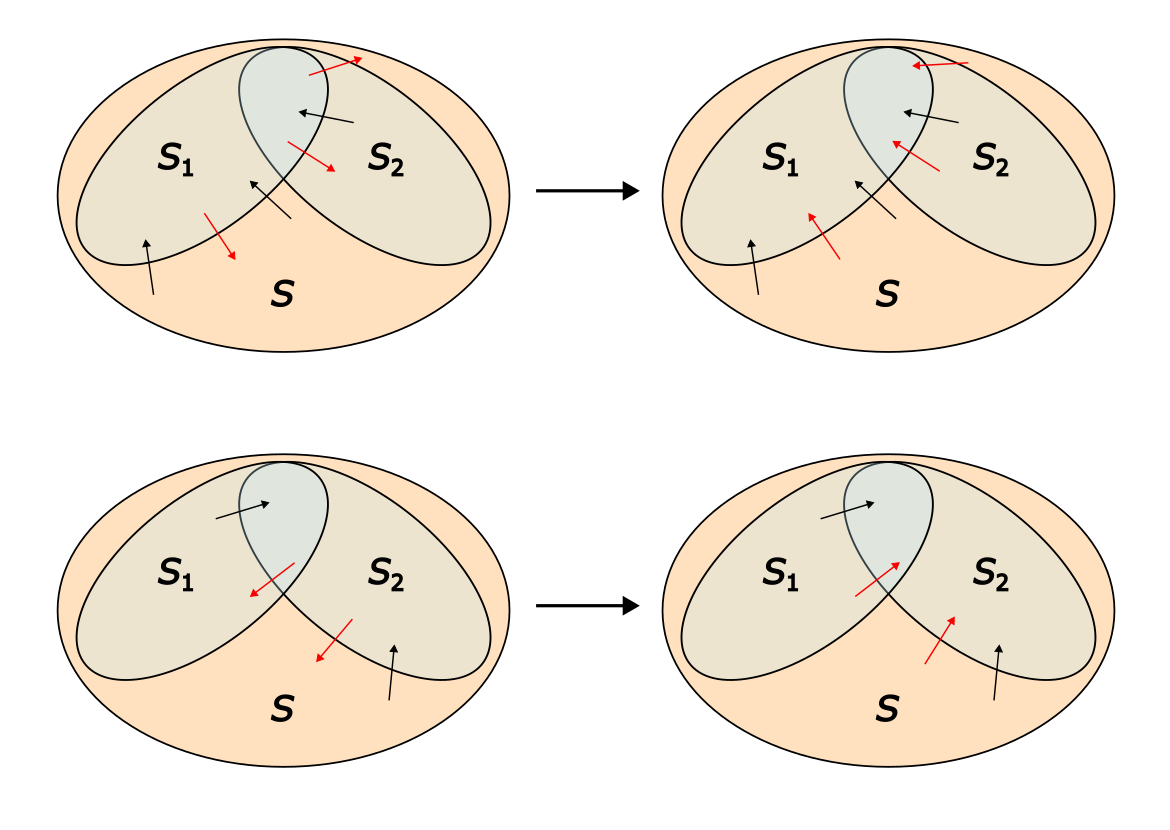}
    \caption{(From the Technical Overview). Our algorithm isolates $S_1$ before recursing on $(S_1,I_1)$, and isolates $S_2$ before recursing on $(S_2,I_2)$. This figure illustrates Steps 2 and 5 in the procedure.}
    \label{fig:steps1_3_2}
\end{figure}
    \begin{enumerate}
        \item For every $v \in S \setminus S_1$, set 
        \[c(v) \coloneq 5\alpha-2  -|E(S_1,v)|.\]
        \begin{enumerate}
            \item Perform the sequence of flips on $\overrightarrow{G}_{i-1}[S \setminus S_1]$ whose existence is guaranteed by \cref{lem:batch} to decrease each $v$'s out-degree to $c(v)$. 
        \end{enumerate}
        \item Orient the edges of $E(S_1,S \setminus S_1)$ so that they point towards $S_1$.
        \item Allow $u_i, ..., u_{j-1}$ to be performed, maintaining the $(5\alpha-2)$-orientation on $\G[S_1]$ whose existence is guaranteed by the inductive hypothesis.
        \item For every $v \in S \setminus S_2$, set 
        \[c(v) \coloneq 5\alpha-2  -|E(S_2,v)|.\]
        \begin{enumerate}
            \item Perform the sequence of flips on $\overrightarrow{G}_{j-1}[S \setminus S_2]$ whose existence is guaranteed by \cref{lem:batch} to decrease each $v$'s out-degree to $c(v)$.
        \end{enumerate}
        \item Orient the edges of $E(S_2, S \sm S_2)$ so that they point towards $S_2$.
        \item Allow $u_{j}, ..., u_{k-1}$ to be performed, maintaining the $(5\alpha-2)$-orientation on $\G[S_2]$ whose existence is guaranteed by the inductive hypothesis.
    \end{enumerate}
See \cref{fig:steps1_3_2} for an illustration of Steps 2 and 5.
\begin{clm}
    The above procedure produces a sequence of $(5\alpha-2)$-orientations.
\end{clm}
\begin{proof}
    Let $v \in S$. We will show that $v$ has out-degree $\le 5\alpha-2$ after each update that occurs during $I$ by showing that no step in the above procedure increases the out-degree of a node to $>5\alpha-2$. We will show this is the case for Steps 1-3 during the interval $I_1$. The proof for Steps 4-6 (during interval $I_2$) is symmetrical by replacing $i$ with $j$, $j$ with $k$, $S_1$ with $S_2$, and $I_1$ with $I_2$.
    \begin{enumerate}
        \item Step 1 sets $c(v) = 5\alpha-2 - |E(S_1,v)| \le 5\alpha-2$ for every $v \in S \sm S_1$. Therefore, the procedure from \cref{lem:batch} will not increase the out-degree of any $v \in S \sm S_1$ to be greater than $5\alpha-2$, and will not affect the out-degrees of nodes in $S_1$.
        \item Step 2 only increases the out-degrees of nodes in $S \sm S_1$. Every node $v \in S \sm S_1$ has its out-degree in $S_1$ increased to $\le |E(S_1,v)|$. However, by the guarantees of \cref{lem:batch}, the out-degree of each $v \in S \sm S_1$ is at most $5\alpha-2 - |E(S_1,v)|$ before Step 2, so Step 2 can only increase a node's out-degree up to $5\alpha-2$.
        \item Step 3 recurses on $S_1$, so it does not increase the out-degrees of any nodes in $S \sm S_1$. Because Step 2 orients every edge in $E(S_1, S\sm S_1)$ toward $S_1$ for the duration of $I_1$, the only edges contributing to out-degrees in $S_1$ during $I_1$ are in $\G[S_1]$. The inductive hypothesis then implies that every node in $S_1$ has out-degree $\le 5\alpha-2$ throughout $I_1$. 
    \end{enumerate}

\end{proof}
The next step is to bound the number of flips the procedure performs. The following lemma will be helpful for us.

\begin{lem} \label{lem:boundary}
    Before Step 1 is performed,
    \[ |\overrightarrow{E}_{G_{i-1}}(S_1,S\setminus S_1)| = O(\Gamma(I)).\]
    Before Step 4 is performed,
     \[ |\overrightarrow{E}_{G_{j-1}}(S_2,S\setminus S_2)| = O(\Gamma(I)).\]
\end{lem}
\begin{proof}
    We will prove the bound on $|\overrightarrow{E}(S_1,S\setminus S_1)|$. The bound on $|\overrightarrow{E}(S_2,S\setminus S_2)|$ is similar. 
    
    Let $e \in \overrightarrow{E}_{G_{i-1}}(S_1,S\setminus S_1)$ before Step 1. There are three cases:
    \begin{enumerate}
        \item $e \in \overrightarrow{E}(A(I),B(I))$,
        \item $e \in \overrightarrow{E}(A(I),X(I))$,
        \item $e$ has tail in $Y(I)$.
    \end{enumerate}
    We bound the number of edges in each case:
    \begin{enumerate}
        \item The number of edges in Case 1 is $|E(A(I),B(I))|$.
        \item The number of edges in Case 2 is $\le (2\alpha-1)|X(I)|$. This is because every node $x\in X(I)$ has $|E(S_1,x)|\le 2\alpha-1$, as $\Hi$ is nearly isolated (see \cref{lem:hier} for details).
        \item The number of edges in Case 3 is $\le (5\alpha-2)|Y(I)|$. This is because $\G[S]$ is assumed to be $(5\alpha-2)$-oriented before Step 1 is performed.
    \end{enumerate}
    Summing over the three cases, we get that
    \[\overrightarrow{E}_{G_{i-1}}(S_1,S\setminus S_1) = O(|E(A(I),B(I))| + \alpha(|X(I)|+|Y(I)|)) = O(\Gamma(I)),\]
    as desired.
\end{proof}
\begin{clm}
    Steps 1 and 4 perform \[O\left(\Gamma(I)\log\left(\frac{\alpha|S|}{\Gamma(I)}\right)\right)\] flips.
\end{clm}
\begin{proof}
    We will show this is the case for Step 1; the proof for Step 4 is symmetric. The bound follows from plugging the parameters of Step 1 into \cref{lem:batch}, with $H \coloneq  S \setminus S_1$ and $\delta \coloneq 5\alpha-2$. Recall that the number of flips performed by \cref{lem:batch} is
    \[O\left(\Phi \cdot \log \left(\frac{(\delta-2\alpha)|V(H)|}{\Phi} \right) \right) = O\left(\Phi \cdot \log \left(\frac{\alpha|S|}{\Phi} \right) \right),\]
    where $\Phi$ is the total excess given as input to the lemma. Our goal then is to bound the total excess $\Phi$ plugged into \cref{lem:batch} as follows:
    \[ \Phi = O( \Gamma(I)).\]
    Recall that each node $v \in S \setminus S_1$ has $c(v) \coloneq 5\alpha-2-|E(S_1,v)|$. Moreover, $v$ has out-degree $\le 5\alpha-2-|\overrightarrow{E}(v,S_1)|$ before Step 1, so the excess of $v$ is
    \[ e(v)\le 5\alpha-2 - |\overrightarrow{E}(v,S_1)| - (5\alpha-2-|E(S_1,v)|) = |\overrightarrow{E}(S_1,v)|.\]
    Therefore, to bound $\Phi = \sum_{v \in S\setminus S_1}e(v)$ it suffices to bound $|\overrightarrow{E}(S_1,S\setminus S_1)|$.
    By \cref{lem:boundary}, 
    \[|\overrightarrow{E}(S_1,S\setminus S_1)| = O(|E(A(I),B(I))| + \alpha|X(I)|+\alpha|Y(I)|) = O(\Gamma(I)),\]
    as desired.
\end{proof}
\begin{clm}
    Steps 2 and 5 perform $O(\Gamma(I))$ flips.
\end{clm}
\begin{proof}
    Note that Step 2 only flips edges that are in $\overrightarrow{E}_{G_{i-1}}(S_1,S \setminus S_1)$ going into Step 2. Because Step 1 does not flip any edges in $\overrightarrow{E}_{G_{i-1}}(S_1,S \setminus S_1)$, these are the same edges that are in $\overrightarrow{E}_{G_{i-1}}(S_1,S \setminus S_1)$ before Step 1. The bound on the number of flips in Step 2 follows then from the first statement of \cref{lem:boundary}.

    Similarly, the bound on the flips in Step 5 follows from the second statement in \cref{lem:boundary}.

\end{proof}

The number of flips performed in Steps 1, 2, 4, and 5 is thus bottle-necked by the number of flips in Steps 1 and 4. 
   \[O\left(\Gamma(I)\log\left(\frac{\alpha|S|}{\Gamma(I)}\right)\right).\]
By the inductive hypothesis, the number of flips performed by Steps 3 and 6 is at most
\[ O\left( \sum_{(S',I') \in desc(S_1,I_1) \cup desc(S_2,I_2)} \Gamma(I')\log\left(\frac{\alpha|S'|}{\Gamma(I')}\right)\right).\]
Thus, the total number of flips is at most
\[ O\left( \sum_{(S',I') \in desc(S,I)} \Gamma(I')\log\left(\frac{\alpha|S'|}{\Gamma(I')}\right)\right),\]
as desired.
\end{proof}
\begin{lem} \label{lem:flips2}
    Let $\G$ be a dynamic graph on node set $V$ undergoing a sequence $U$ of arboricity $\alpha$-preserving edge insertions and deletions, such that $|U|$ is a power of $2$. Let $(S,I)$ be the root of the special DIS hierarchy defined on $\G$ during $U$, constructed as in \cref{lem:hier}. Suppose there is a $(5\alpha-2)$-orientation $\overrightarrow{G}_0$ on $G_0$. There is a sequence of
    \[ O\left( \alpha |U|\log\log(|U|) \right)\]
    flips so that after each update every $v \in S$ has out-degree $\le 5\alpha-2$ in $S$.
\end{lem}
\begin{proof}
    By \cref{lem:flips}, there is a sequence of
    \[ O\left( \sum_{(S,I) \in V^{int}(\Hi)}  \Gamma(I)\log\left(\frac{\alpha|S|}{\Gamma(I)}\right)\right)\]
    flips maintaining a $(5\alpha-2)$-orientation within $S$. By the log sum inequality (see \cref{apx:log_sum}),
    \begin{align*}
         \sum_{(S,I) \in V^{int}(\Hi)}  \Gamma(I)\log\left(\frac{\alpha|S|}{\Gamma(I)}\right) \le \left( \sum_{(S,I) \in V^{int}(\Hi)}\Gamma(I)\right)\log\left(\frac{\sum_{(S,I) \in V^{int}(\Hi)}\alpha|S|}{\sum_{(S,I) \in V^{int}(\Hi)}\Gamma(I)}\right),
    \end{align*}
    so our goal now is to bound the RHS of this inequality. Recall that we have the following bounds from \cref{lem:hier}:
    \begin{enumerate}
        \item $\sum_{(S,I) \in V^{int}(\Hi)} |X(I)| = O(|U|)$
        \item $\sum_{(S,I) \in V^{int}(\Hi)} |Y(I)| = O(|U|)$
        \item $\sum_{(S,I) \in V(\Hi)} |S| = O(|U|\log|U|)$
        \item $\sum_{(S,I) \in V^{int}(\Hi)} |E(A(I),B(I))|=O( \alpha|U|)$.
    \end{enumerate}
    By \cref{def:gamma},
    \[ \Gamma(I) = \alpha |X(I)| +\alpha|Y(I)| + |E(A(I),B(I))|,\]
    so by using bounds 1, 2, and 4 above, we have that
    \[\sum_{(S,I) \in V^{int}(\Hi)} \Gamma(I)= O(\alpha|U|).\]
    Combining this fact with bound 3, we have that
    \begin{align*}
         \left( \sum_{(S,I) \in V^{int}(\Hi)}\Gamma(I)\right)\log\left(\frac{\sum_{(S,I) \in V^{int}(\Hi)}\alpha|S|}{\sum_{(S,I) \in V^{int}(\Hi)}\Gamma(I)}\right) &= O\left(\alpha|U|\log\left(\frac{\alpha|U|\log|U|}{\alpha|U|}\right)\right) \\
         &= O(\alpha|U|\log \log |U|),
    \end{align*}
    as desired.

\end{proof}
Finally, we can prove \cref{thm:main}.
\main*
Before we begin the proof, observe that the theorem follows immediately from \cref{lem:flips2} so long as we can safely assume the following two facts without loss of generality:
\begin{enumerate}
    \item $|U|$ is a power of 2.
    \item $|U| = \text{poly}(n)$,
\end{enumerate}
where we need the second fact in order to show that $\alpha|U|\log \log |U| = O(\alpha|U| \log \log n)$. The proof below works by partitioning $U$ into intervals of length $p$, where we define $p$ to be the power of 2 between $n/2$ and $n$, and applying \cref{lem:flips2} to each interval individually.
\begin{proof}[Proof of \cref{thm:main}.]
    First, we can assume without loss of generality that $|U| \ge n-1$. Else, $V$ can be partitioned into $S_1$, $S_2$ such that no edge ever exists crossing $(S_1,S_2)$, so we can consider the update sequences restricted to $S_1$ and $S_2$ individually.
    
    Next, let $p \in \Z$ be the power of 2 in $[n/2,n-1]$. We can assume without loss of generality that $|U|$ is a multiple of $p$. Else, add $p-(|U|\text{ mod }p)$ arbitrary updates to the end of $U$. Because $|U| \ge n-1 \ge p-1$, this increases $|U|$ by at most a factor of $2$.

    Finally, subdivide $U$ into intervals $I_1, ..., I_q$ where every interval contains $p$ updates. For each interval $I_r$, construct the set $S_r \subseteq V$ as follows:
    \begin{enumerate}
        \item Let $S_r$ be the set of nodes incident to at least one edge added or deleted during $I_r$.
        \item While there is a node $v \in V \sm S_r$ with $|E(S_r,v)|\ge 2\alpha$ throughout $I_r$:
        \begin{enumerate}
            \item Add $v$ to $S_r$.
        \end{enumerate}
    \end{enumerate}
    Let $\Hi_r$ be the special hierarchy defined on $\G[S_r]$ throughout $I_r$, constructed as in \cref{lem:hier}. Note that $(S_r,I_r)$ is the root of $\Hi_r$. The desired flip sequence will be as follows:
    \begin{enumerate}
        \item For each $I_r \in \{I_1, ..., I_q \}$:
        \begin{enumerate}
            \item Let $[i,j) = I_r$.
            \item 
            Obtain an $\alpha$-orientation on $G_{i-1}[V\sm S_r]$ by 1-orienting each forest in $G_{i-1}[V\sm S_r]$'s forest decomposition.
            \item Orient every edge in $E(S_r, V\sm S_r)$ toward $S_r$.
            \item Allow the interval $I_r$ to proceed, using $\Hi_r$ to maintain the $(5\alpha-2)$-orientation on $S_r$ whose existence is guaranteed by \cref{lem:flips2}.
        \end{enumerate}
    \end{enumerate}
    Our goal now is to show that the above procedure maintains a $(5\alpha-2)$-orientation. 
    
    First, we show that every $v\in V\sm S_r$ has out-degree $\le 3\alpha-1$ at every time step of $I_r$. Step (b) ensures that $v$ has out-degree $\le \alpha$ in $V \sm S_r$, and by the near isolation of $(S_r,I_r)$, $|E(v,S_r)| \le 2\alpha-1$. Therefore, $d^+(v) \le 3\alpha-1$ after Steps (b) and (c), and the totality of $(S_r,I_r)$ means that $v$ is not incident to any edge inserted during $I_r$.
    
    Second, note that each $v\in S_r$ has out-degree $\le 5\alpha-2$, as Step (c) ensures that $v$ has no out-neighbors in $V\sm S_r$, and \cref{lem:flips2} ensures that it only ever has out-degree $\le (5\alpha-2)$ in $S_r$.

    Finally, let us bound the number of edge flips required. For each interval $I_r$, Steps (b) and (c) together reorient $\le \alpha n = O(\alpha p)$ edges, as they reorient every edge in $G$ at most once. Across $q$ intervals, Steps (b) and (c) therefore perform $O(\alpha pq) = O(\alpha|U|)$ edge flips. 

    Thus, the number of required flips is dominated by Step (d). By \cref{lem:flips2}, in each interval $I_r$, Step (d) performs $O(\alpha p\log\log p) = O(\alpha p \log \log n)$ flips. Summing over the $q$ intervals, this amounts to $O(\alpha \cdot p\cdot q\cdot\log\log n) = O(|U|\cdot \alpha\log\log n)$ flips.
\end{proof}

\paragraph{Acknowledgments.} 

Nicole Wein thanks her collaborators from when she first worked on this problem 10 years ago: Josh Alman, Shay Solomon, and Virginia Vassilevska Williams. She is especially grateful to Shay Solomon for many extended discussions.

More recently, Nicole Wein thanks the participants of the open problem sessions of her \emph{Graph Algorithms} course for discussions about the problem, in particular  Teo Miklethun, Rahul Ramesh, Rayen Tan, Michael Yi Wang, and especially Daniel Skora. She also thanks Erik Demaine for use of his ``supercollaboration'' method and accompanying \emph{Coauthor} software, for the open problem sessions.

\paragraph{AI Acknowledgment.}
The authors originally obtained a proof of the main theorem in the case $\alpha =1$ without use of generative AI outside of literature review. GPT 5.5 Plus was used only to generalize the earlier version of \cref{lem:batch} from the arboricity-$1$ case to the general arboricity case. We have verified and rewritten the proof for clarity.

\bibliographystyle{plain}
\bibliography{ref}

\appendix

\section{Log sum inequality} \label{apx:log_sum}
The log sum inequality is stated as follows \cite{ct05}.
\begin{thm}
    Let $a_1, ..., a_n$ and $b_1, ..., b_n$ be sequences of non-negative real numbers, and let $a = \sum_{i=1}^n a_i$ and $b = \sum_{i=1}^n b_i$. Then
    \[ \sum_{i=1}^n a_i\log\frac{a_i}{b_i}\ge a \log\frac{a}{b}.\]
\end{thm}
We reverse the sign of each side by flipping the numerator and denominator inside the logarithms:
\[ \sum_{i=1}^n a_i\log\frac{b_i}{a_i}\le a \log\frac{b}{a}.\]
This is the version used in the proof of \cref{lem:flips2}.

\end{document}